\documentclass{article}
\usepackage[preprint]{spconf, graphicx}
\usepackage[usenames, dvipsnames]{color}
\usepackage[colorlinks=true, allcolors=blue]{hyperref}
\usepackage{epsfig,amssymb,amsmath}
\usepackage{calc}
\usepackage{ragged2e}
\usepackage{array}
\usepackage{bm}

\newcommand{\T}{^\intercal}
\newcommand{\R}{\mathbb{R}}

\newcommand{\eg}{\textit{e}.\textit{g}. }
\newcommand{\ie}{\textit{i}.\textit{e}., }

\renewcommand{\vec}[1]{\bm{#1}}

\usepackage[hang,flushmargin]{footmisc}

\title{Who do I sound like?\\Showcasing Speaker Recognition Technology by YouTube Voice Search}

\name{Ville Vestman, Bilal Soomro, Anssi Kanervisto, Ville Hautam{\"a}ki, Tomi Kinnunen\thanks{This research was partially funded by the Academy of Finland (grants \#313970 and \#309629).}}
\address{School of Computing, University of Eastern Finland}

\begin{document}

\ninept

\setlength{\abovedisplayskip}{4pt plus 1.0pt minus 1.0pt}
\setlength{\belowdisplayskip}{4pt plus 1.0pt minus 1.0pt}
\maketitle
\begin{abstract}
\vspace{-1mm}
The popularization of science can often be disregarded by scientists as it may be challenging to put highly sophisticated research into words that general public can understand. This work aims to help presenting speaker recognition research to public by proposing a publicly appealing concept for showcasing recognition systems. We leverage data from YouTube and use it in a large-scale voice search web application that finds the celebrity voices that best match to the user's voice. The concept was tested in a public event as well as ``in the wild'' and the received feedback was mostly positive. The i-vector based speaker identification back end was found to be fast (665 ms per request) and had a high identification accuracy (93\%) for the YouTube target speakers. To help other researchers to develop the idea further, we share the source codes of the web platform used for the demo at \url{https://github.com/bilalsoomro/speech-demo-platform}.
\end{abstract}

\vspace{-3mm}
\section{Introduction}
\vspace{-2mm}

As methodology researchers, we often find it challenging to explain intuitively where and how our research advancements in speaker recognition can be used. To demonstrate speaker recognition technology in an appealing way to the public, many challenges need to be resolved. Besides the standard challenges of speaker recognition technology such as background noise \cite{zhao2014robust}, channel mismatch~\cite{Solomonoff2005}, and the requirement of fast response times in large-scale recognition tasks \cite{Schmidt2014}, there are challenges related to the demo design itself. First, the traditional speaker recognition setting requires at least two separate speech inputs from the user, one is for enrollment and the other one for test. The requirement of two separate recordings can be inconvenient for an user who wants to quickly test the system. The second challenge in showcasing is how to give an attractive feedback to the user. This could be implemented as a real-life application, for example, by using user's voice to open a physical lock, or in a less involved way by displaying recognition scores in a screen \cite{lee2011joint}.

\begin{figure}[t!]

  \centerline{\includegraphics[width=0.8\linewidth]{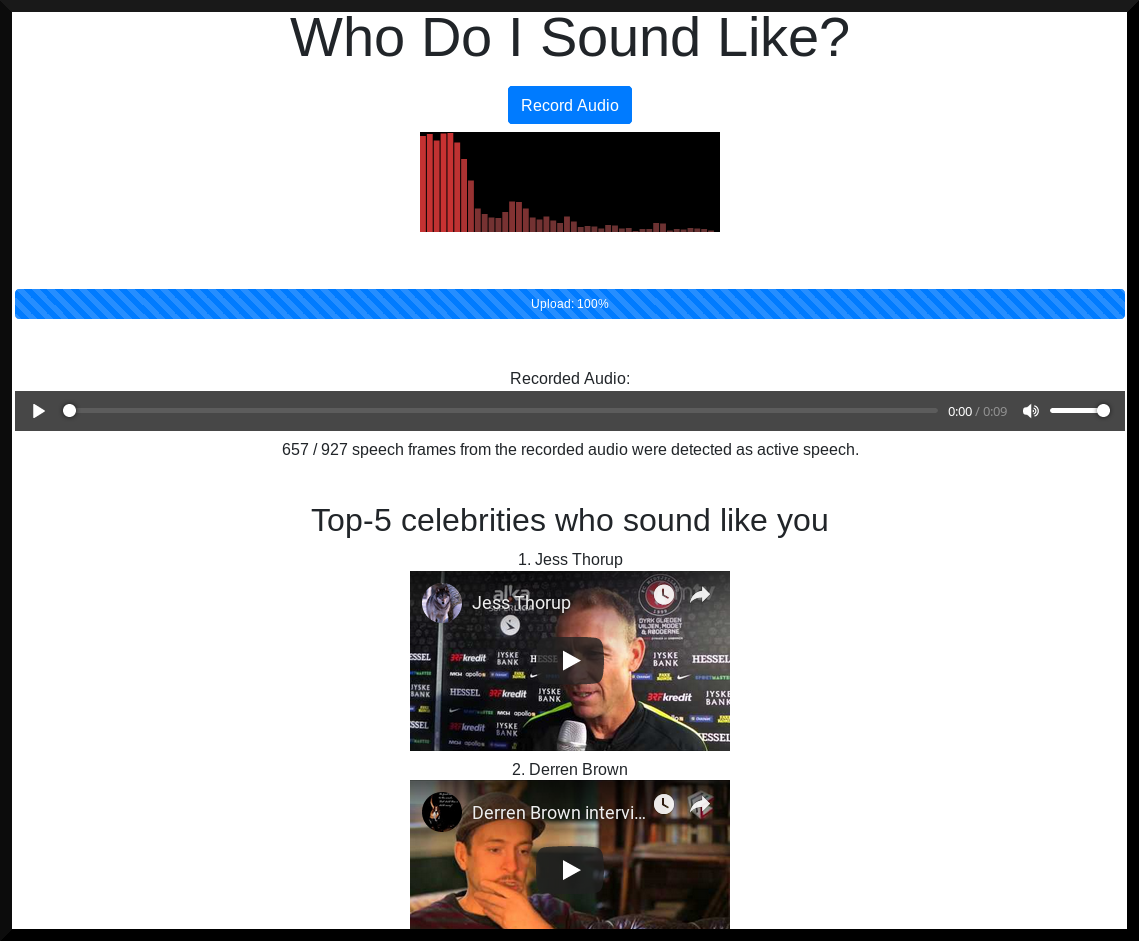}}
    \vspace{-2mm}
    
    \caption{A screenshot from our voice search web application displaying the basic elements of the UI: Recording button, audio visualization, playback option for the recorded speech, and the results.}
    \label{fig:screenshot}
\end{figure}

In this work, we present a concept for creating publicly appealing demos to showcase speaker recognition technology by leveraging public-domain target speaker data collected from YouTube. The core idea is to compare users' speech to the ones of celebrities on YouTube, who have been enrolled prior to the real-time demonstration. The results of the comparison are then displayed as a selection of YouTube videos from the best matching celebrities, which allows users to see and listen to the celebrity speakers who they most resemble to (Figure \ref{fig:screenshot}). Even if we focus on speaker recognition research, the same concept could also be applied for other things that can be inferred or estimated from speech such as age, emotion, or language. For example, if the user records angry voice, the results could show YouTube videos of angry people. When the results include famous public figures, the user's interest and satisfaction of the demo tends to naturally rise. We saw this positive effect while presenting the demo in a locally organized sub-event of an European-wide ``The European Researchers' Night 2018'' \footnote{\url{https://ec.europa.eu/research/mariecurieactions/actions/european-researchers-night}} that aims to bring scientific research to public.

We run our demo on a web platform that can be used on PCs and mobile phones with an internet connection to ensure good accessibility of the demo. The web platform communicates with a computation server that runs the speaker recognition back end based on our recent work on computationally efficient i-vector extraction \cite{Vestman2018}. The back end provides the results to the web platform that displays them by using embedded YouTube video players.

Our extensive use of YouTube data has been made possible by recent automated speech data collection efforts in \cite{Nagrani2017} and \cite{Chung2018} resulting in VoxCeleb1 and VoxCeleb2 corpora, respectively. These corpora provide a large set of annotated YouTube speech data including metadata for obtaining web links to the original YouTube videos.

To best of our knowledge, prior existing speaker recognition demos have not utilized VoxCeleb data in the proposed way. We are aware of a website \footnote{\url{https://celebsoundalike.com}} with a similar idea, but unfortunately we have not been able to successfully run the demo to see how it functions. Based on the celebrity speaker names, that demo does not utilize VoxCeleb data, and likely does not display YouTube videos in the results.

In summary, the current work describes a novel concept that allows speech technology research teams to demonstrate their research without requiring large amount of additional work. To help other researchers to apply the concept for their own research, we share the source code of our web platform allowing a quick start for prototyping possible demo applications. We tested the concept among the public using our voice comparison demo utilizing standard speaker recognition techniques, and the received feedback from the people was mainly positive.

\label{sec:intro}

\vspace{-2mm}
\section{Web platform for speech demos}
\vspace{-2mm}
\label{sec:web-platform}

We designed the web platform with the sole purpose of demonstrating speech processing systems to the public, and in this work we used it to demonstrate speaker recognition using YouTube data. The platform is implemented as a web service in PHP and JavaScript, supporting different browser and devices. Users can select one of speech processing ``methods'' defined by the host of the platform. The methods could, for example, perform speaker recognition or age estimation from the recorded speech. 

The back end of the platform is implemented in PHP, and thus only needs a web server capable of running PHP (\eg Apache or Nginx). For simplicity and reliability, server-side code only receives WAV audio files from the clients and runs a specified method as a \texttt{system()} call, and finally returns the results to the client. For privacy reasons, the audio file is not stored on the server, and is immediately removed at the end of handling user's request. The platform supports including additional user inputs required for the analysis, \eg the claimed identity for speaker verification demo.

The front end of the platform is implemented in JavaScript, which also handles recording of the audio in raw format at 16kHz. Features required by this code are supported by the most PCs and Android phones, making it easier to share the demo with others. The user interface records a sample of user's speech, queries what speech processing method should be applied on the recorded sample, sends the sample to server to be processed, and displays the results. 

We share the source code of the platform in the hopes it will support other researchers in speech analysis to demonstrate their work to the public. The code includes instructions how to setup the server in couple of steps. New speech analysis systems for demonstration can be added by modifying a single JSON file.

\vspace{-2mm}
\section{Speaker recognition back end}
\vspace{-2mm}
\label{sec:speaker-recognition}

The system comparing user's voice to voices in YouTube videos can be regarded as a \emph{closed set speaker identification} system. As we only utilize a closed set of YouTube target speakers, we can include the data from the target speakers in the system development. In this section, we describe the data sets and the speaker identification system used for providing functionality to the web front end.

\vspace{-2mm}
\subsection{YouTube data: VoxCeleb1 \& VoxCeleb2}
\vspace{-1mm}
The audio-visual VoxCeleb corpora \cite{Nagrani2017, Chung2018} have been adopted in many application areas including speaker recognition \cite{Nagrani2017, Chung2018}, speech separation \cite{Afouras2018}, and emotion recognition \cite{Albanie18a} to name a few. The VoxCeleb data has been automatically collected from YouTube by exploiting face verification and active speaker detection systems. An automated pipeline enabled collecting very large scale speaker recognition data sets: When combined, the VoxCeleb corpora consist of almost 1.3 million speech clips from over 170,000 YouTube videos from more than 7000 speakers and, in total, nearly 3000 hours of speech material. The average length of speech clips in VoxCeleb is about eight seconds.

The metadata provided with the VoxCeleb corpora includes, for example, speakers' names, IDs of the original YouTube videos, and the starting and ending times of the clips within the videos expressed as frames. This metadata is enough for setting up a demo where users can find best matching voices to theirs from YouTube. Although the metadata is automatically obtained, it is, in our experience, fairly accurate. Regarding to the correctness of the labels, the authors of VoxCeleb mention that the VoxCeleb2 corpus is mainly intended to be used as a training data set and that during the data collection thresholds for discarding false positives were not as strictly set as with VoxCeleb1 data collection \cite{Chung2018}. We have witnessed a few labeling errors in VoxCeleb2, such as Finnish president Tarja Halonen being confused to talk-show host Conan O'Brien. However, the errors do not exist to an extent that would be a considerable problem for our application.

\vspace{-2mm}
\subsection{Speaker identification system description}
\vspace{-2mm}

The acoustic feature vectors of the speaker identification system consist of 20 MFCCs plus their delta and double-delta coefficients. 
The system discards non-speech frames using a energy based speech activity detector and normalizes obtained features to have zero mean and unit variance.

For training the system components and enrolling the celebrities, we used those speakers from VoxCeleb corpora who had more than five utterances of length of five seconds or more. There are 903,498 such utterances and 7,363 such speakers. In the training of some system components, only a fraction of this data was needed to reach close to optimal recognition accuracy. We trained an universal background model (UBM) using one-thirtieth of the selected 903,498 utterances. The UBM is a 1024-component Gaussian mixture model (GMM) \cite{reynolds2000speaker}, which is used to compute sufficient statistics for i-vector extraction. We compute 800-dimensional i-vectors by compressing mean supervectors of maximum a posteriori (MAP) adapted GMMs using probabilistic principal component analysis (PPCA) as described in \cite{Vestman2018}. This is a (speed-wise) high-performing alternative to the stardard i-vector extraction that is traditionally done via front-end factor analysis \cite{dehak2011front, kenny2012small}. We trained the PPCA model using one-fifteenth of the selected data.

Prior to scoring, i-vectors are centered using the mean computed from the whole training data of 903,498 utterances and then normalized to unit length. Scoring is performed with a simplified Gaussian probabilistic linear discriminant analysis (G-PLDA) model \cite{garcia2011analysis}, which has a 350-dimensional speaker subspace. The G-PLDA model was trained using the whole training data.

At the online stage, the i-vector extracted from user's recording is scored against all of the 903,498 i-vectors used in PLDA training. The speakers are sorted according to the scores of their highest scoring utterances, from highest score to lowest. Finally, the system sends the names of the \mbox{top-5} speakers together with the links to the YouTube-videos that correspond to the highest scoring utterances to the client.

\renewcommand{\arraystretch}{1.3}
\begin{table*}[t!]
\caption{Speaker rank testing for six public figures using 10 audio clips from each speaker. The speaker ranks range from 1 to 5 and 'x' is shown if the result list of top-5 speakers did not contain the correct speaker at all. The tests are performed with and without a replay channel. The replay experiment does not require direct access to the back end system, but can be done by using the web demo only.}
\vspace{2mm}
\label{table:rankings}
\small
\begin{tabular}{p{0.20\linewidth-2\tabcolsep} p{0.20\linewidth-2\tabcolsep}>{\raggedleft\arraybackslash} p{0.05\linewidth-2\tabcolsep}>{\raggedleft\arraybackslash} p{0.05\linewidth-2\tabcolsep}>{\raggedleft\arraybackslash} p{0.05\linewidth-2\tabcolsep} p{0.05\linewidth-2\tabcolsep} p{0.20\linewidth-2\tabcolsep}>{\raggedleft\arraybackslash}p{0.05\linewidth-2\tabcolsep}>{\raggedleft\arraybackslash} p{0.05\linewidth-2\tabcolsep}>{\raggedleft\arraybackslash}p{0.05\linewidth-2\tabcolsep}}
& \multicolumn{4}{l}{Without replay channel} & & \multicolumn{4}{l}{With replay channel} \\
\cline{2-5} \cline{7-10}
& & \multicolumn{3}{l}{Occurrences in} & & & \multicolumn{3}{l}{Occurrences in} \\
\cline{3-5} \cline{8-10}
Speaker's name & list positions for 10 clips & top1 & top3 & top5 & & list positions for 10 clips & top1 & top3 & top5 \\
\hline
Hillary Clinton & \texttt{1111111111} & 10 & 10 & 10 & & \texttt{1111111111} & 10 & 10 & 10\\
Ariana Grande & \texttt{1111111111} & 10 & 10 & 10 & & \texttt{3111111111} & 9 & 10 & 10\\
Oprah Winfrey & \texttt{1111111111} & 10 & 10 & 10 & & \texttt{1311111112} & 8 & 10 & 10\\
Johnny Depp & \texttt{1111112112} & 8 & 10 & 10 & & \texttt{1111x11121} & 8 & 9 & 9\\
Bruno Mars & \texttt{1411111112} & 8 & 9 & 10 & & \texttt{1x21111211} & 7 & 9 & 9\\
Conan O'Brien & \texttt{1111111111} & 10 & 10 & 10 & & \texttt{1111111111} & 10 & 10 & 10\\
\hline
Total (in \% of max.)&  & 93 & 98 & 100 & & & 87 & 97 & 97 \\
\hline
\end{tabular}
\end{table*}
\renewcommand{\arraystretch}{1.0}

\vspace{-1mm}
\subsection{System runtime considerations at online stage}
\vspace{-1mm}

To ensure fast response times, we implemented the speaker recognition back end as a server that has all the necessary models preloaded in the memory. The server is implemented with Python using scientific computing libraries available to it (\eg NumPy and SciPy). We pay special attention to the PLDA scoring and i-vector extraction as they are the most time consuming steps during the computation.

In \cite{garcia2011analysis}, it is shown that the score for a trial using G-PLDA can be computed as
\[
    \textrm{score} = \vec{\tilde\eta}_1 \T \widetilde Q \vec{\tilde\eta}_1 + \vec{\tilde\eta}_2 \T \widetilde Q \vec{\tilde\eta}_2 + 2 \vec{\tilde\eta}_1 \T \Lambda \vec{\tilde\eta}_2 + \textrm{const},
\]
where $\vec{\tilde\eta}_1$ and $\vec{\tilde\eta}_2$ are lower dimensional projections of enrollment and test i-vectors, respectively, and where $\vec{\tilde\eta}_1 \T \widetilde Q \vec{\tilde\eta}_1$ and $\vec{\tilde\eta}_1 \T \Lambda$ can be precomputed.

As we work with an identification system (one test segment vs. all enrollment segments), the second term $ \vec{\tilde\eta}_2 \T \widetilde Q \vec{\tilde\eta}_2$ is a constant and thus can be neglected. Therefore, to get all the $n=903{,}498$ scores at online stage, we only need to compute
\[
    \textrm{scores} = \vec{\nu} + 2 D P \vec{\eta}_2,
\]
where $\vec{\nu}$ is an $n$-dimensional vector containing precomputed values $\vec{\tilde\eta}_1 \T \widetilde Q \vec{\tilde\eta}_1$, matrix $D \in \R^{n \times 350}$ contains precomputed vectors $\vec{\tilde\eta}_1 \T \Lambda$, and $P$ is a $350 \times 800$ projection matrix that projects test i-vector $\vec{\eta}_2$ to a lower dimensional space so that $\vec{\tilde\eta}_2 = P \vec{\eta}_2$. The product $D\vec{\tilde\eta}_2$ can be efficiently parallelized.

The i-vector extraction using PPCA is simply a matter of compressing 61440-dimensional GMM-supervector to 800-dimensional space using a precomputed projection matrix. Note that the traditional approach for i-vector extraction would, in addition, require inverting an $800 \times 800$ posterior covariance matrix~\cite{madikeri2014fast, Vestman2018}. 

\vspace{-1mm}
\section{System evaluation}
\vspace{-2mm}

We tested our voice search demo and the underlying speaker recognition back end in multiple ways using both objective and subjective measures in evaluation. On the objective side, we computed an equal error rate (EER) using VoxCeleb speaker verification protocol and further we tested the rankings that the system displays for newly downloaded and replayed YouTube data. On the subjective side, we gathered feedback from the users of the system, including their opinions on how close the displayed top five celebrities sound to the user. 

\vspace{-1mm}
\subsection{Evaluation using VoxCeleb speaker verification protocol}

\vspace{-1mm}
The VoxCeleb1 speaker verification test protocol includes 37720 trials with a balanced number of same speaker trials and impostor trials. The trial list has been formed using 4715 utterances from 40 speakers. Using this protocol, we obtained EER of 6.69 \%. This result is better than the baseline result for i-vectors in \cite{Chung2018}, but should not be directly compared as our system utilizes testing utterances also in system training.

\vspace{-1mm}
\subsection{Speaker rank testing on non-VoxCeleb YouTube data}
\vspace{-1mm}
To test the the final deployed demo, we studied the speaker rankings the system outputs. For this purpose, we collected a small set of new YouTube data. This set contains 10 new speech clips for six public figures in VoxCeleb corpora. The clips are about 15 seconds long each and they are extracted from videos that are not already present in VoxCeleb corpora. When the new clips are fed to the speaker recognition back end, the output lists of top-5 speakers should contain the correct speaker as they are present in VoxCeleb and hence are already enrolled to the system.

The new test data was used with the system in two ways: First, we downloaded the speech clips from YouTube and fed the data directly to the speaker recognition back end. Secondly, we played files directly from YouTube and at the same time recorded them with the web demo. Unlike the first approach, the second one includes the channel effects caused by replaying the data. In the replay experiment, the playback device was Sony SRS-XB10 portable Bluetooth speaker while the web demo was ran in Chrome browser in Nokia 8 smartphone running Android $8{.}1{.}0$. The distance between the two devices was kept to 5 cm as the recording device was held by hand above the up facing speaker. The room in which the experiment took place was quiet and the only background noise that was present was the fan noise of the laptop which was connected to the speaker.

For both settings, with and without replay, the speaker rankings for all the test utterances are shown in Table \ref{table:rankings}. In addition, the table contains statistics of the number of occurrences in the top-1, top-3, and top-5 rankings. Without the replay, the system was always able to include the correct speaker to the top-5 list and 93\% of the times the speaker was identified correctly (\ie in top-1). Replaying the audio clips decreased the system performance only slightly as the correct speaker was left outside the top-5 list only twice out of the 60 trials.

\begin{figure}[t!]

  \centerline{\includegraphics[width=0.9\linewidth]{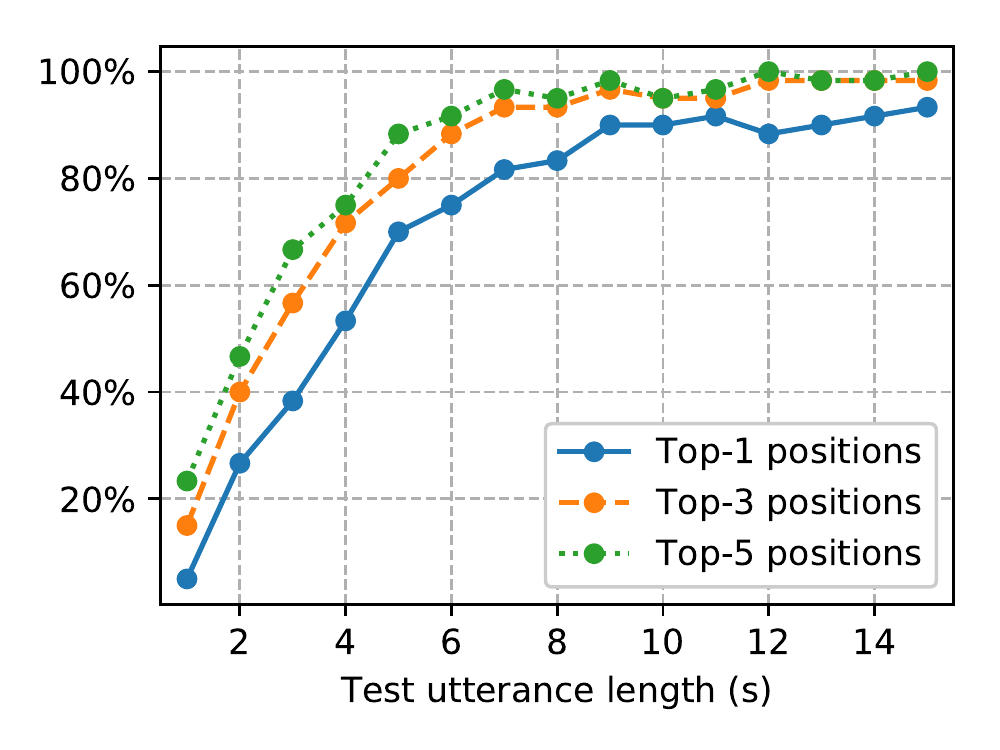}}
    
    \vspace{-5mm}
    \caption{The effect of utterance length on speaker ranking performance. Specifically, the graph shows how often the target speakers are displayed in the top-lists when tested with different lengths of test utterances from the target. An utterance of length 9s is required to reach close to optimal performance.}
    \label{fig:length}
\end{figure}

To get insight of how long of an utterance is required for getting good results in our celebrity matching demo, we studied the effect of length of the test utterance on system accuracy. We ran the previous experiment without the replay effect using utterances clipped to lengths ranging from 1 second to 15 seconds. We found that the test segment needs to be at least 9 seconds to obtain close to optimal performance and at least 5 seconds to obtain identification accuracies greater than 70\% (Figure \ref{fig:length}).

\renewcommand{\arraystretch}{1.3}
\begin{table}[t!]
\caption{Computation times for the different steps in the voice comparison pipeline. The steps marked with \textbf{*} are parallelized to 16 CPU cores while others steps utilize only 1 core. The total response time is the time it takes to upload the speech and compute and display the results. The data was collected from 402 requests, except for the total response time which was collected together with the feedback questionnaire (n=27).}
\vspace{-2mm}
\label{table:times}
\small
\begin{tabular}{p{0.55\linewidth-2\tabcolsep}>{\raggedleft\arraybackslash}p{0.15\linewidth-2\tabcolsep}>{\raggedleft\arraybackslash}p{0.15\linewidth-2\tabcolsep}>{\raggedleft\arraybackslash}p{0.15\linewidth-2\tabcolsep}}
& \multicolumn{3}{l}{Times in milliseconds (ms)} \\
\cline{2-4}
& median & mean & SD \\
\hline
Audio loading, MFCC extraction & 47.1 & 86.2 & 142.1 \\
Sufficient statistics computation & 20.9 & 46.3 & 98.1 \\
MAP adaptation & 0.8 & 0.9 & 0.6 \\
\mbox{Supervector compression (PPCA)\textbf{*}} & 42.6 & 56.5 & 28.3 \\
I-vector centering \& length norm. & 0.1 & 0.1 & $<0.1$ \\
I-vector compression (PLDA) & 0.4 & 0.5 & 0.3 \\
PLDA scoring\textbf{*} & 336.4 & 423.8 & 195.8 \\
Sorting speakers & 39.6 & 44.6 & 13.8 \\
\hline
Total time in computing server & 521.5 & 661.0 & 331.6 \\
\hline
Total response time & 1791.1 & 2503.5 & 1975.1 \\
\hline
\end{tabular}
\end{table}
\renewcommand{\arraystretch}{1.0}

\vspace{-2mm}
\subsection{Feedback and impressions from public testing}
\vspace{-2mm}
The first public test for our voice search demo took place in the event ``The European Researchers' Night 2018'' (September 28, 2018), where researcher's from many fields were displaying their research to the public. The event was funded by EU and it was organized in many countries across the Europe. In our local event, we were showcasing our demo for five hours and for the most of the time there was a long queue of people waiting for their turn to test our demo. In total, approximately 150 people tried the demo. The feedback was mostly positive, although not everyone was satisfied with their results. As the event was targeted for families, many of the testers were children. This was a slight problem as only a small minority of the speakers in VoxCeleb corpora are children, causing it to be difficult to find a good voice match for everyone.

In the event, we were using our own high-quality microphone (Zoom H6 Handy Recorder, XY mic) and a laptop that was well tested with the demo. To see how the demo works ``in the wild'', we shared a web link to our demo in a multiple social media platforms. The shared demo application was equipped with a short feedback questionnaire for subjective evaluation. We also collected error reports containing system information of the devices on which the demo did not work.

\begin{figure}[tb!]

  \centerline{\includegraphics[width=\linewidth]{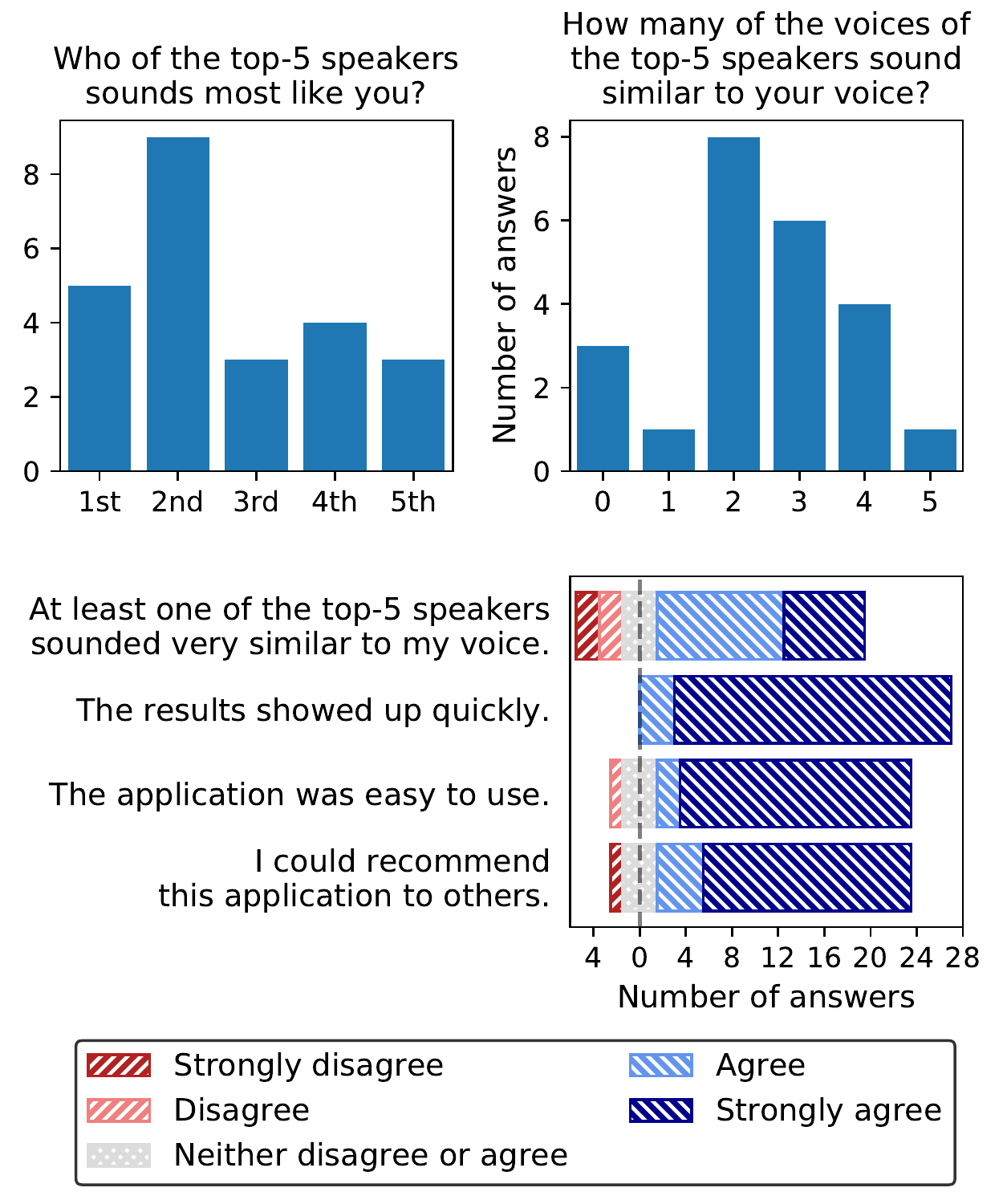}}
    \vspace{-3mm}
    \caption{Results from the feedback questionnaire, gathered from users using platform that finds matches for their speech from a set of over 7000 celebrities. Based on these subjective assessments, the system is able to find good matches for users' speech in most cases.}
    \label{fig:questionnaire}
\end{figure}

The public testing revealed that the device and browser support is still quite limited due to some issues with the audio recording and playback support. Based on the feedback, we estimate that demo ran on 50 to 70 percent of the device-browser configurations that our test users were using. We also got some good suggestions how to improve the user interface and we believe that together with improved browser support the user experience can be very good as the received answers (n=27) to the questionnaire were already fairly positive as can be seen from Figure \ref{fig:questionnaire}.

\vspace{-2mm}
\subsection{Response and computation times}
\vspace{-1mm}

During the test in the wild, we collected computation times of the different steps in the voice comparison. The statistics are summarized in Table \ref{table:times}. The average time to compute one voice comparison request was 661 milliseconds, which means that our computation server could, theoretically, respond to 5000 requests in an hour without processing multiple requests in parallel. The total response time, on average, was about 2.5 seconds. As seen from Figure \ref{fig:questionnaire}, this level of responding speed was considered to be fast.

\vspace{-2mm}
\section{CONCLUSIONS}
\vspace{-2mm}
We successfully capitalized the appeal to public figures with our YouTube voice search demo application. The objective and the subjective evaluations of the demo showed that the platform was mostly successful in providing good results and also being convenient to use. The feedback received from the users allows us to further develop our demo platform, which we have shared for open source development at \url{https://github.com/bilalsoomro/speech-demo-platform}. We would be happy to see the proposed concept to be applied in the future with other speech related recognition systems as well.



\vfill\pagebreak

\bibliographystyle{IEEEbib}
\bibliography{refs}

\end{document}